\documentclass[preprintnumbers,aps,prd,reprint,groupedaddress,nofootinbib]{revtex4-1}
\pdfoutput=1

\usepackage{booktabs}
\usepackage{footnote}

\usepackage{lipsum}
\usepackage{amsmath}    
\usepackage{amssymb}    
\usepackage{color}         
\usepackage{graphicx}     
\usepackage[T1]{fontenc}
\usepackage{times}         
\usepackage{mathrsfs} 
\usepackage{myterms}
\usepackage{hyperref} 
\hypersetup{
    colorlinks=true,       
    linkcolor=blue,        
    citecolor=blue,        
    filecolor=magenta,     
    urlcolor=blue          
}
\usepackage[all]{hypcap} 

\usepackage[none]{hyphenat} 

\usepackage{color,xcolor,fancybox,epsf,rotating,colordvi}

\begin{document}

\preprint{\parbox{1.6in}{\noindent arXiv:2003.01662\\WHU-HEP-PH-TEV007}}

\title{Funnel annihilations of light dark matter and the invisible decay 
of the Higgs boson}

\author{Kun Wang}
\email[]{wk2016@whu.edu.cn}
\affiliation{Center for Theoretical Physics, School of Physics and Technology, Wuhan University, Wuhan 430072, China}

\author{Jingya Zhu}
\email[]{zhujy@whu.edu.cn} 
\affiliation{Center for Theoretical Physics, School of Physics and Technology, Wuhan University, Wuhan 430072, China}

\date{\today}

\begin{abstract}
The semi-constrained NMSSM (scNMSSM), or NMSSM with non-universal Higgs masses, can naturally predict a light dark matter under current constraints including Higgs data, sparticle-mass bounds, dark matter searches, and muon g-2, etc.
In this work, we take this scenario of scNMSSM as an example to study the funnel-annihilation mechanisms of light dark matter ($1\!\!\thicksim\!\!62\GeV$) and the invisible Higgs decay.
In this scenario we found that:
(i) There can be four funnel-annihilation mechanisms for the LSP $\tilde{\chi}^0_1$, which are the $h_2$, $Z$, $h_1$ and $a_1$ funnel.
(ii) For the $h_1$ and $a_1$ funnel with right relic density, the $\tilde{\chi}^0_1$ mass is lighter than $12\GeV$, and the invisible Higgs decay can be $2\%$ at most.
(iii) For the $h_2$ and $Z$ funnel with right relic density, the invisible Higgs decay can be about $0.4\%$ and $1\%$ respectively at most.
(iv) If the invisible Higgs decay was discovered at the HL-LHC, the four funnel-annihilation mechanisms of light dark matter may be all excluded with $\tilde{\chi}^0_1$ as the only dark matter source.
Four benchmark points, one for each mechanism, are proposed for future checking with updated experimental results.
\end{abstract}


\maketitle

\section{Introduction}
\label{sec:intro}

Dark Matter (DM) is one of the most fascinating mysteries in the universe.
Particle dark matter is strongly supported by overwhelming evidence in astrophysical observations \cite{Jungman:1995df, Bertone:2004pz, Bertone:2016nfn}.
Among many dark matter candidates suggested by theorists, the weakly interacting massive particle (WIMP), also known as the thermal dark matter, is an influential one, whose mass can be $\Ocal(1)\MeV \!\sim\! \Ocal(100)\TeV$ \cite{Boehm:2002yz, Boehm:2003bt, Murayama:2009nj, Hambye:2009fg} and with which the DM relic density is naturally interpreted through the so-called freeze-out mechanism \cite{Bernstein:1985th, Srednicki:1988ce}, and also the history of the Big Band Nucleosynthesis (BBN) and recombination in the early universe are successfully interpreted.
However, present negative DM search results, especially these from direct detection experiments, seriously erode the surviving space of WIMP at the electroweak scale.
Thus, it motivates us to focus on lighter DM, especially its annihilation mechanisms that can not be checked by direct searches.

Correlations between Higgs and DM have been widely studied for several tens of years \cite{Shrock:1982kd, Griest:1987qv, Gunion:1993jf, Choudhury:1993hv, Eboli:2000ze, Belanger:2001am, Bento:2001yk, Belotsky:2002ym, Godbole:2003it, Davoudiasl:2004aj, Accomando:2006ga, Draper:2010ew, Cai:2011kb, Cao:2011re, He:2011de, Pospelov:2011yp}, especially after the first hint of the $125\GeV$ Higgs at the end of 2011 \cite{Lebedev:2011iq, Baek:2011aa, Bai:2011wz, He:2011gc, Djouadi:2011aa, Baer:2011ab, Drozd:2011aa, Kadastik:2011aa, Wang:2012ts, Cao:2012im, Djouadi:2012zc, LopezHonorez:2012kv, Ellis:2012aa, Joglekar:2012vc, Cline:2012hg, Baer:2012uya, Baek:2012se, Espinosa:2012vu, Belanger:2013kya, Goudelis:2013uca, Carpenter:2013xra, Wang:2013yba, Han:2013ic, Khoze:2013uia, Banerjee:2013fga, Baek:2014jga, Ko:2014gha, Feng:2014vea, Berlin:2014cfa, Hamaguchi:2015rxa, Krnjaic:2015mbs, Escudero:2016gzx, Goncalves:2016bkl, Banerjee:2016nzb, vanBeekveld:2016hug, Barman:2017swy, Englert:2017aqb, Yin:2017wxm, Biekotter:2017gyu, Baum:2017enm, Han:2018bni, Ahriche:2018ger, Held:2018cxd, Harz:2019rro, Chanda:2019xyl, Arcadi:2019lka, Heisig:2019vcj, Wang:2019jhb, vanBeekveld:2019tqp, Kato:2020pyl, Pozzo:2018anw, Han:2016gyy, Cao:2015efs}.
If Higgs has interaction with DM and DM mass is lighter than half of the Higgs mass, the invisible Higgs decay offers another way to detect DM indirectly.
Recently according to the Run-I and Run-II data at the LHC, the upper limit of the invisible Higgs decay reaches $26\%$ by ATLAS \cite{Aaboud:2019rtt}, and $19\%$ by CMS \cite{Sirunyan:2018owy}.
While the future accuracy for that can reaches to $5.6\%$, $0.5\%$, $0.24\%$ and $0.26\%$ according to the future detections HL-LHC \cite{Gomez-Ceballos:2013zzn}, FCC \cite{dEnterria:2017dac}, CEPC \cite{Tan:2020fxk,Liu:2016zki} and ILC \cite{Ishikawa:2019uda} respectively.
So there is much space to study the nature of light DM from the invisible Higgs decay, which we discuss in this work.

Supersymmetry (SUSY) can both predict a natural SM-like Higgs and a light DM candidate, and can also ensure the unification of three gauge interactions, thus has attracted much attention.
There are various concrete models and scenarios in the framework of SUSY, among which a scenario in the semi-constrained NMSSM (scNMSSM) can survive with the above advantages under current constraints including Higgs data, sparticle-mass bounds, DM searches, and muon g-2, etc \cite{Wang:2020tap}.
In this work, we take this scenario in the scNMSSM as an example, to discuss the annihilation mechanisms of light DM from the invisible Higgs decay.

The remainder of this paper is organized as follows.
In \sref{sec:ana} we introduce the scNMSSM and relevant analytic calculations briefly.
In \sref{sec:num} we present the numerical calculations and discussions.
Finally, we draw our main conclusions in \sref{sec:conc}.

\section{The model and analytic calculations}
\label{sec:ana}
The NMSSM extends the MSSM by a singlet superfield $\hat{S}$, which provides an effective $\mu$ term when $\hat{S}$ gets a VEV.
The superpotential of the $\mathbb{Z}_3$-invariant NMSSM is
\begin{equation}
    W_{\rm NMSSM} = W_{\rm Y} + \lambda \hat{S} \hat{H}_u \!\!\cdot\!\! \hat{H}_d + \frac{\kappa}{3} \hat{S} ^3
\end{equation}
where the Yukawa term $W_{\rm Y}$ is the same as that of MSSM, the $\hat{H}_{u,d}$ denote the doublet superfields, and the $\lambda$, $\kappa$ are the dimensionless couplings.
After electroweak symmetry breaking, the doublet and singlet scalar fields mix to generate 3 CP-even Higgses $h_{1,2,3}$ and 2 CP-odd Higgses $a_{1,2}$ respectively, and their superpartners $\tilde{H}_{u,d}^0$ (higgsinos) and $\tilde{S}$ (singlino) mix with gauginos $\tilde{W}^0$ and $\tilde{B}$, generating 5 neutralinos $\tilde{\chi}^0_i$ ($i=1,2,3,4,5$).
Hereafter we use $\chi$ to denote the lightest neutralino (LSP) $\tilde{\chi}^0_1$ for  convenience.

The scNMSSM is also called the NMSSM with non-universal Higgs masses (NUHM), for it assumes the unification of gauginos and sfermions respectively at the GUT scale, while not that of the Higgs sector \cite{Wang:2020tap, Wang:2019biy, Wang:2018vxp, Ellwanger:2014dfa, Das:2013ta, Ellwanger:2018zxt, Ellwanger:2016sur}.
In the scNMSSM, the gauginos are constrained to be very heavy, for the high mass bound of gluino and the unification of gauginos at the GUT scale.
And because of the large interactions with the SM sector, the higgsino-dominated LSP usually predicts very small relic density.
Thus the lightest neutralino $\chi$ predicting right relic density is usually singlino-dominated in the scNMSSM.
If singlino is lighter than the SM-like Higgs, the singlet-dominated scalar is usually also lighter than $125\GeV$, thus $h_1,\, a_1$ are usually singlet-dominated scalars, and $h_2$ is usually the SM-like Higgs with $m_{h_2}\!\simeq\!125\GeV$.
Considering $\chi$ as mixed only by singlino and higgsinos, we can write the couplings between Higgs and $\chi$ as \cite{Ellwanger:2004xm, Maniatis:2009re}
\begin{eqnarray}
C_{h_a \chi\chi}
&=& \sqrt{2}\lambda (S_{a1} N_{12}N_{13} +
S_{a2} N_{11}N_{13} + S_{a3} N_{11}N_{12}) \quad \nonumber \\
&& - \sqrt{2} \kappa S_{a3}N_{13}N_{13} \,,
\end{eqnarray}
\begin{eqnarray}
C_{a_a \chi \chi} &= i&\left[\sqrt{2}\lambda (P_{a1}
N_{12}N_{13} + P_{a2} N_{11}N_{13} + P_{a3} N_{11}N_{12}) \right. \nonumber \\
&& \left. - \sqrt{2}\kappa P_{a3} N_{13} N_{13}\right] \,,
\end{eqnarray}
where $N_{1i}$ ($i=1,2,3$) are the coefficients of $\tilde{H}_u$, $\tilde{H}_d$ and $\tilde{S}$ in $\chi$ respectively.
Similarly, $S_{ai}$ and $P_{ai}$ are the coefficients of $H_u$, $H_d$ and $S$ in $h_a$ and $a_a$ respectively.
When $\chi$ is singlino-dominated, $h_2$ is SM-like, $h_1$ and $a_1$ are singlet-dominated,
\begin{eqnarray}
&& C_{h_1\chi\chi} \approx -\sqrt{2}\kappa,
\qquad\quad
C_{a_1\chi\chi} \approx -i\sqrt{2}\kappa,\nonumber\\
&& C_{h_2\chi\chi} \approx \sqrt{2}\lambda N_{11} \!-\! \sqrt{2}\kappa S_{23} .
\end{eqnarray}

When $2m_{\chi}\!<\!m_{\phi}$, with $\phi=h_2,h_1,a_1$, the invisible width of the SM-like Higgs $h_2$ can be calculated as 
\begin{eqnarray}
\Gamma_{\phi\to\chi\chi}=\frac{C_{\phi\chi\chi}^2}{16\pi} m_{\phi} \left(1-\frac{4m^2_{\chi}}{m^2_{\phi}}\right)^{3/2} \,.
\end{eqnarray}

To calculate the relic density of the LSP $\chi$ with velocity $v$, one has to solve firstly the number density $n$ from the Boltzmann equation
\begin{eqnarray}
 \frac{dn}{dt} = -3 H n - \langle v \sigma_{\rm eff}\rangle
 \left( n^2 - n_{\rm eq}^2 \right) \,,
\end{eqnarray}
where $H$ is the Hubble rate, $n_{\rm eq}$ is the density in thermal equilibrium,  $\sigma_{\rm eff}$ is the total cross scetion of dark matter annihilation or co-annihilation, and $\langle v \sigma_{\rm eff}\rangle$ is the thermal average of $v \sigma_{\rm eff}$.
For the s-channel annihilation processes, two dark matter particles annihilate to a medium boson $\phi$, and $\phi$ decay promptly into two SM particles, which is also called $\phi$-funnel annihilation \cite{Jungman:1995df}.
According to perturbation theory, the cross section of  $\phi$-funnel annihilation to a pair of fermions $f\bar{f}$ at leading order (LO) can be written as \cite{Hamaguchi:2015rxa}
\begin{eqnarray}
 \sigma_{\rm eff}^{\phi,f\bar{f}} &=& \frac{1}{2}C_{\phi\chi\chi}^2 \frac{s\Gamma_{\phi\to f\bar{f}}/m_{\phi}}{(s-m_\phi^2)^2+m_\phi^2 \Gamma_\phi^2} \sqrt{1-\frac{4m_{\chi}^2}{s}}
\end{eqnarray}
where 
$s$ is the squared total center-of-mass energy,
$\Gamma_{\phi}$ and $\Gamma_{\phi\to f\bar{f}}$ are the total width and partial $f\bar{f}$ width of $\phi$ decay respectively.
The total annihilation cross section $\sigma_{\rm eff}$ is the sum to all channels including different mediums particles and different final states.
For cold dark matter, whose average velocity is non-relativistic, one can expand $v\sigma_{\rm eff}$ in powers of $v^2$, obtaining 
\begin{eqnarray}
v\sigma_{\rm eff} \equiv a + b v^2 + {\cal O}\left( v^4 \right)
\end{eqnarray}
Then the relic density can be expressed as \cite{Young:2016ala}
\begin{eqnarray}
\Omega_\chi h^2
&\equiv& m_\chi n_0 \frac{h^2}{\rho_c}
\nonumber\\
&=&\left(\frac{m_{\chi}/T_{\rm f}}{10}\right)\sqrt{\frac{g_*}{100}} ~ \frac{0.847 \!\times\! 10^{-27}{\rm cm}^3{\rm s}^{-1}}{\langle v \sigma_{\rm eff}\rangle}
\end{eqnarray}
where $n_0$ is the present number density, $g_*$ is the total effective degrees of freedom, $T_{\rm f}$ is the freezing-out temperature, and $\rho_{c}\!\!=\!\!3H_0^2/(8\pi G_N)$ is the critical density of the universe.

The calculation of the DM relic density in SUSY models are performed detailly in several public codes, e.g., the {\sf micrOMEGAs} \cite{Belanger:2001fz, Belanger:2005kh, Gunion:2005rw}, where all relevant annihilation and co-annihilation processes are implemented.

\section{Numerical calculations and discussions}
\label{sec:num}

\begin{figure*}[htbp]
\centering
\includegraphics[width=.9\textwidth]{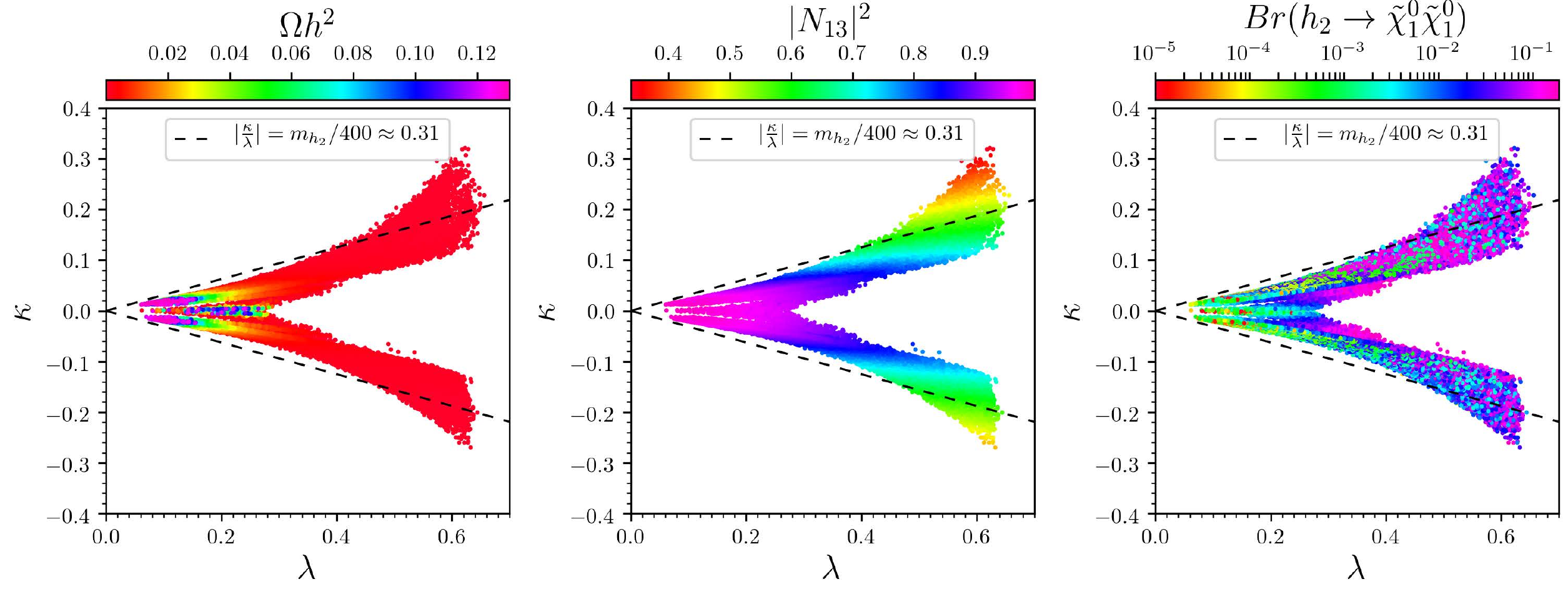}\qquad\qquad
\vspace{-0.4cm}
\caption{(color online). The surviving samples projected in the $m_{a_1}$ versus $m_{\chi}$ planes, with colors indicating the relic density $\Omega h^2$ (left), the singlino component $|N_{13}|^2$ of the LSP $\chi$ (middle) and Higgs invisible branching ratio $Br(h_2\!\!\to\!\!\chi\chi)$ (right) respectively.
The dashed line is $\displaystyle |\kappa/\lambda|= 125/400 \approx 0.31$.
The $\chi$ denotes the lightest neutralino $\tilde{\chi}^0_1$ in this paper.
}
\label{fig1}
\end{figure*}

\begin{figure*}[!htbp]
\centering
\includegraphics[width=.6\textwidth]{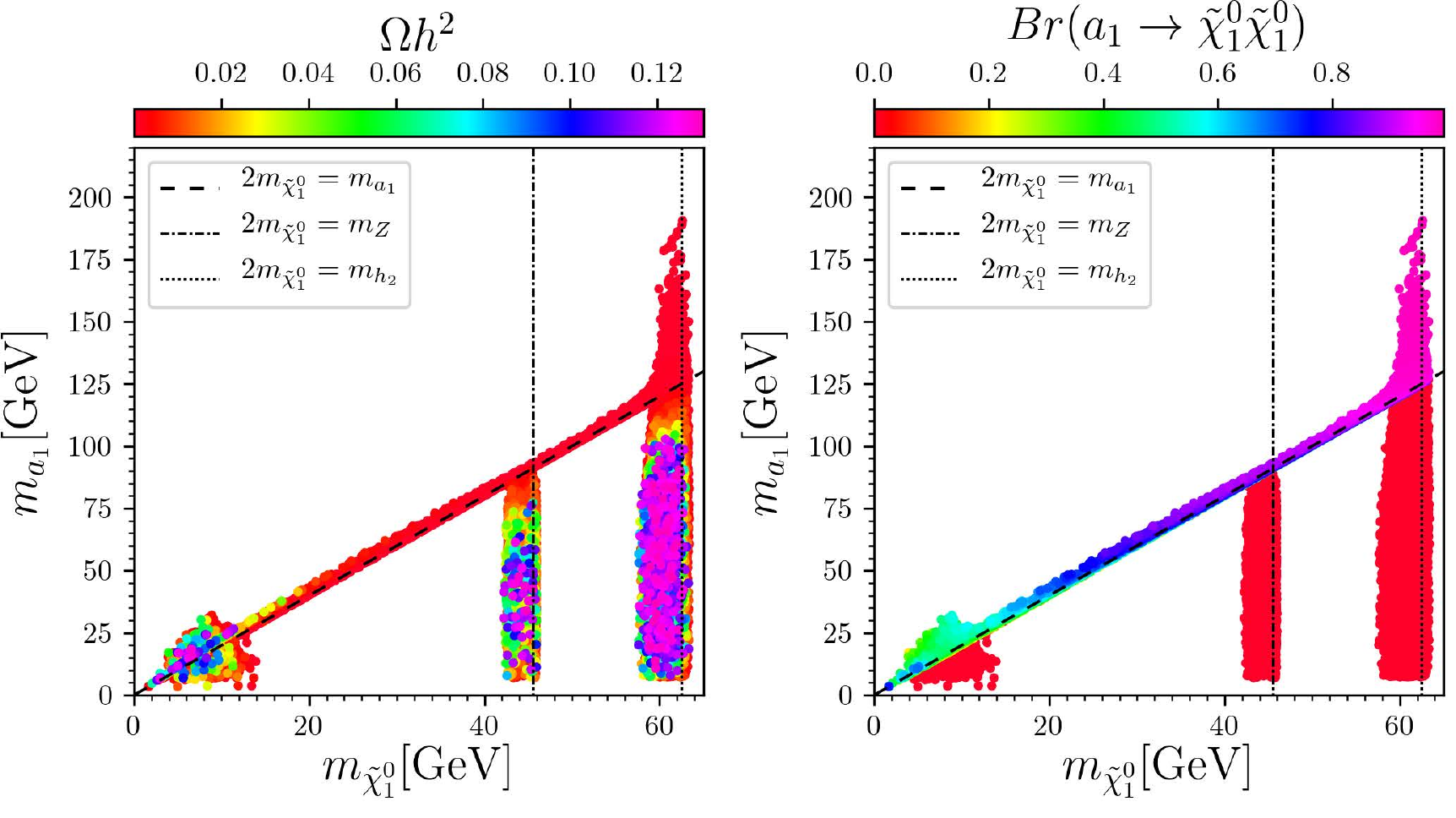}
\includegraphics[width=.36\textwidth]{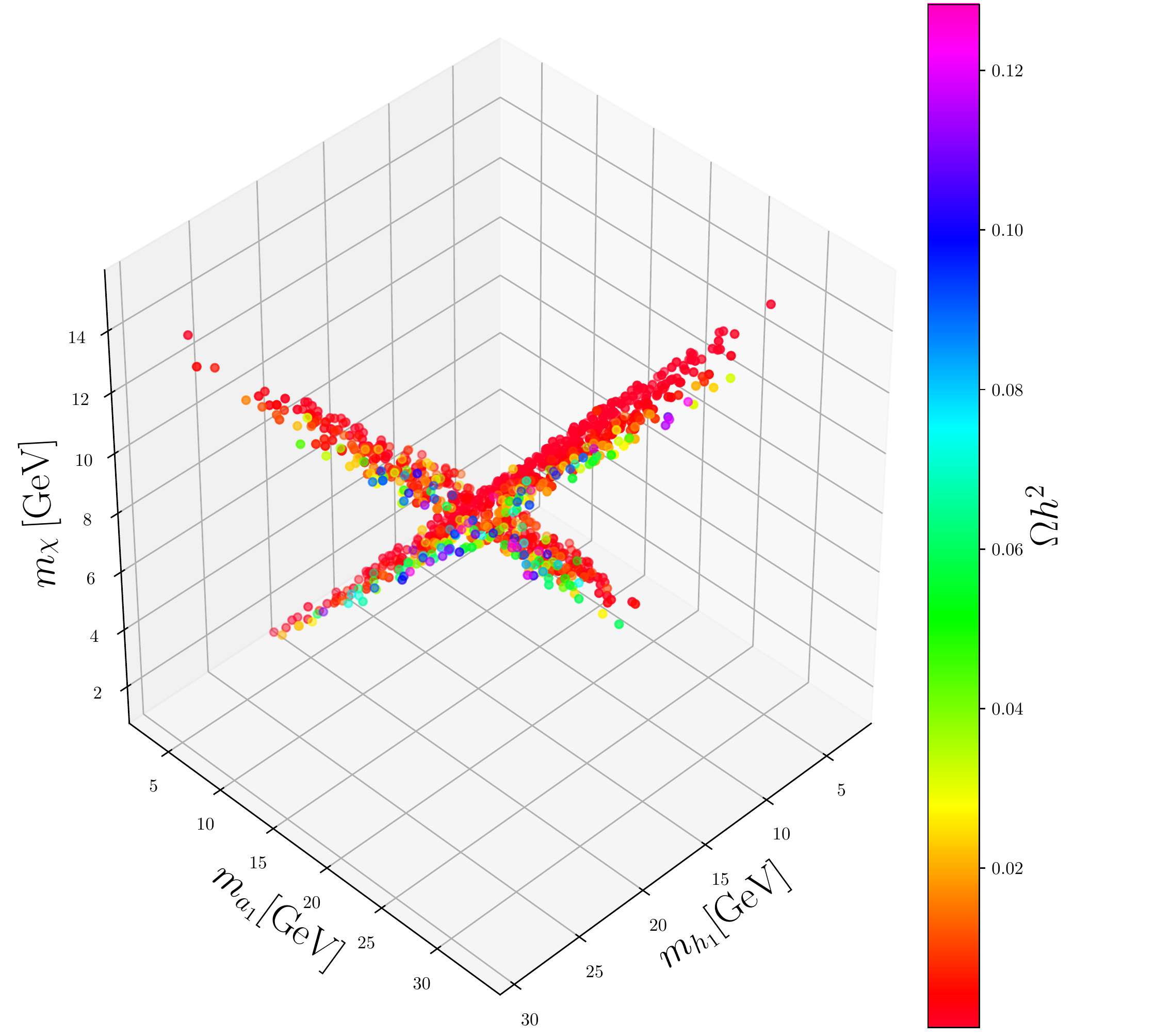}
\vspace{-0.4cm}
\caption{(color online). The surviving samples projected in the $m_{a_1}$ versus $m_{\chi}$ planes (left and middle), with colors indicating the relic density $\Omega h^2$ (left and right) and the invisible branching ratio $Br(a_1\!\!\to\!\! \chi \chi)$ (middle) respectively.
The dashed, dash-doted and doted lines denote $2m_{\chi}=m_{a_1}, m_Z {~\rm and~} 125\GeV$ respectively.
In the right panel, only samples with $m_{\chi}\!\!\leqslant\!\! 15\GeV$ are projected, and the three axis are $m_{a_1}$, $m_{h_1}$ and $m_{\chi}$ respectively.
The $\chi$ denotes the lightest neutralino $\tilde{\chi}^0_1$ in this paper.
}
\label{fig2}
\end{figure*}

In this work, we use the result we got in our former work on scNMSSM \cite{Wang:2020tap}, where we used \textsf{NMSSMTools-5.5.2}
\cite{Ellwanger:2004xm, Ellwanger:2005dv, Ellwanger:2006rn} to generate the particle spectrum, using {\sf micrOMEGAs-5.0} \cite{Belanger:2001fz, Belanger:2005kh, Gunion:2005rw} there to calculate the DM relic density and $\chi$-nucleon cross section, imposing the corresponding constrains there, and also used \textsf{HiggsBounds-5.5.0} \cite{Bechtle:2015pma,Bechtle:2013wla,Bechtle:2011sb,Bechtle:2008jh} to constraint the extra Higgses, used \textsf{SModelS-v1.2.2} \cite{Kraml:2013mwa, Ambrogi:2017neo, Ambrogi:2018ujg, Dutta:2018ioj, Buckley:2013jua, Sjostrand:2006za, Sjostrand:2014zea} with official 1.2.2 database to constrain the sparticle sectors.
Thus eventually the surviving samples we are using here satisfy the constraints including Higgs data, muon g-2, B physics, sparticle searches, DM relic density and direct searches, etc.
As shown in Ref.\cite{Wang:2020tap}, the particle spectrum of this scenario is characterized by
\begin{itemize}
  \item In the colored sectors, the gluginos and the first two generation squarks are heavier than 2 TeV, thus $M_{1/2}$ at the GUT scale, or $M_3/2.4\simeq M_2/0.8 \simeq M_1/0.4$ at the $M_{\rm SUSY}$ scale due to RGE runnings, is lager than about $1\TeV$. And due to the RGE runnings including $M_3$, even the lightest squark, $\tilde{t}_1$, is also over $700\GeV$.
  \item In the slepton sector, due to the RGE runnings including $M_2$ and $M_1$, the charged and uncharged sleptons in the first two generations are heavier than about $300\GeV$, while only $\tilde{\tau}_1$ and $\tilde{\nu}_\tau$  in the third generation can be lighter and to about $100\GeV$.
  \item In the chargino sector, $\tilde{\chi}^\pm_2$ are wino-like and heavier than about $800\GeV$, while $\tilde{\chi}^\pm_1$ are higgsino-like at $100\!\!\sim\!\!200 \GeV$.
  \item In the neutralino sector, $\tilde{\chi}^0_5$ is wino-like and heavier than $800\GeV$, $\tilde{\chi}^0_4$ is bino-like and heavier than $400\GeV$, the two higgsino-like neutralinos are $100\!\!\sim\!\!200 \GeV$, while the singlino-like neutralino can be $1\!\!\sim\!\!120 \GeV$.
  \item In the Higgs sector, $h_2$ is the SM-like Higgs with mass at $123\!\!\sim\!\!127 \GeV$, $h_1$ is singlet-dominated and lighter than $123\GeV$, while the light CP-odd Higgs $a_1$ is also singlet-dominated but can be lighter or heavier than $m_{h_2}$.
      The invisible Higgs decay caused by $m_{\chi}\!\!\lesssim\!\! 62\GeV$ is lower than $19\%$ as constrained by the LHC result \cite{Aaboud:2019rtt, Sirunyan:2018owy}, the Higgs exotic decays caused by $m_{h_1,a_1}\lesssim62\GeV$ is also lower than about $20\%$ for $h_2$ is SM-like.
\end{itemize}
In the following, we focus on the lightest neutralino (LSP) $\chi$ with mass at $1\!\!\sim\!\!62\GeV$ in the scNMSSM: its annihilation mechanisms and relic density, and the invisible Higgs decay into them.

In Fig.\ref{fig1} we project the surviving samples in the $\kappa$ versus $\lambda$ planes, with colors indicate the relic density $\Omega h^2$ and singlino component $|N_{13}|^2$ of the LSP $\chi$ respectively.
From Fig.\ref{fig1}, we can see that
\begin{itemize}
  \item For samples with the right relic density, $\lambda\lesssim0.3$, $|\kappa/\lambda|\lesssim0.3$, $\chi$ is highly singlino-dominated, and the invisible Higgs decay $Br(h_2\!\!\to\!\!\chi\chi)\lesssim0.1$.
  \item For samples with $\lambda\gtrsim0.3$, there can be large mixing between higgsino and singlino in $\chi$, and the singlino component is inversely proportional to $|\kappa/\lambda|$, while that of higgsino is opposite.
  \item For samples with $\lambda\gtrsim0.3$, the relic density of $\chi$ is smaller than $0.02$, for the large interaction between singlino and SM-like Higgs caused by large $\lambda$ and/or sizable higgsino component.
      And also for the same reason, most samples have sizable invisible branching ratio $Br(h_2\!\!\to\!\!\chi\chi)\gtrsim0.1$. Besides, we checked that for most of the samples, the singlet component in $h_2$ $|S_{23}|^2 \!\ll\! 0.1$.
\end{itemize}

In Fig.\ref{fig2} we project the surviving samples in the $m_{a_1}$ versus $m_{\chi}$ planes, with colors indicate the relic density $\Omega h^2$ and the invisible branching ratio $Br(a_1\to \chi \chi)$ respectively.
According to this figure, we can sort the surviving samples into four cases, and combine with Fig.\ref{fig1} we can see that:
\begin{itemize}
  \item {\bf Case I:} $2 m_{\chi}\!\thickapprox\! m_{h_2}$, thus the annihilation mechanism of $\chi$ is $h_2$ funnel, or $h_{\rm SM}$ funnel. This case mainly locates in the $0.05\!\lesssim\!\lambda\!\lesssim\!0.3$ and $|\kappa/\lambda|\!<\! 0.3$ region.
  \item {\bf Case II:} $2 m_{\chi}\!\thickapprox\! m_{Z}$, thus the annihilation mechanism of $\chi$ is $Z$ funnel. This case also mainly locates in the $0.05\!\lesssim\!\lambda\!\lesssim\!0.3$ and $|\kappa/\lambda|\!<\! 0.3$ region, but with smaller $m_{\chi} \!\simeq\! 2\kappa \mu/\lambda$.
  \item {\bf Case III:} $2 m_{\chi}\!\thickapprox\! m_{a_1}$, thus the annihilation mechanism of $\chi$ is $a_1$ funnel. This case can locate in all the surviving region in Fig.\ref{fig1}, with $m_{\chi}$ varying in $1\!\!\sim\!\!62\GeV$, while only in $1\!\!\sim\!\!12\GeV$ the $\chi$ can provide right relic density.
  \item {\bf Case IV:} $2 m_{\chi}\!\thickapprox\! m_{h_1}$, thus $\chi$ is $h_1$ funnel. This case can only locate in the $0.05\!\lesssim\!\lambda\!\lesssim\!0.3$ and $|\kappa/\lambda|\!\ll\! 0.3$ region, with $m_{\chi}\!\lesssim\! 15\GeV$.
\end{itemize}
From the middle panel, we can also see that when $2 m_{\chi}< m_{a_1}$, the invisible branching ratio $a_1\to\chi\chi$ is proportional to $m_{\chi}$, for $\chi$ and $a_1$ are both from the singlet superfield, and their interaction and $\chi$ mass are both proportional to the parameter $\kappa$.

\begin{figure}[!htbp]
\centering
\includegraphics[width=.36\textwidth]{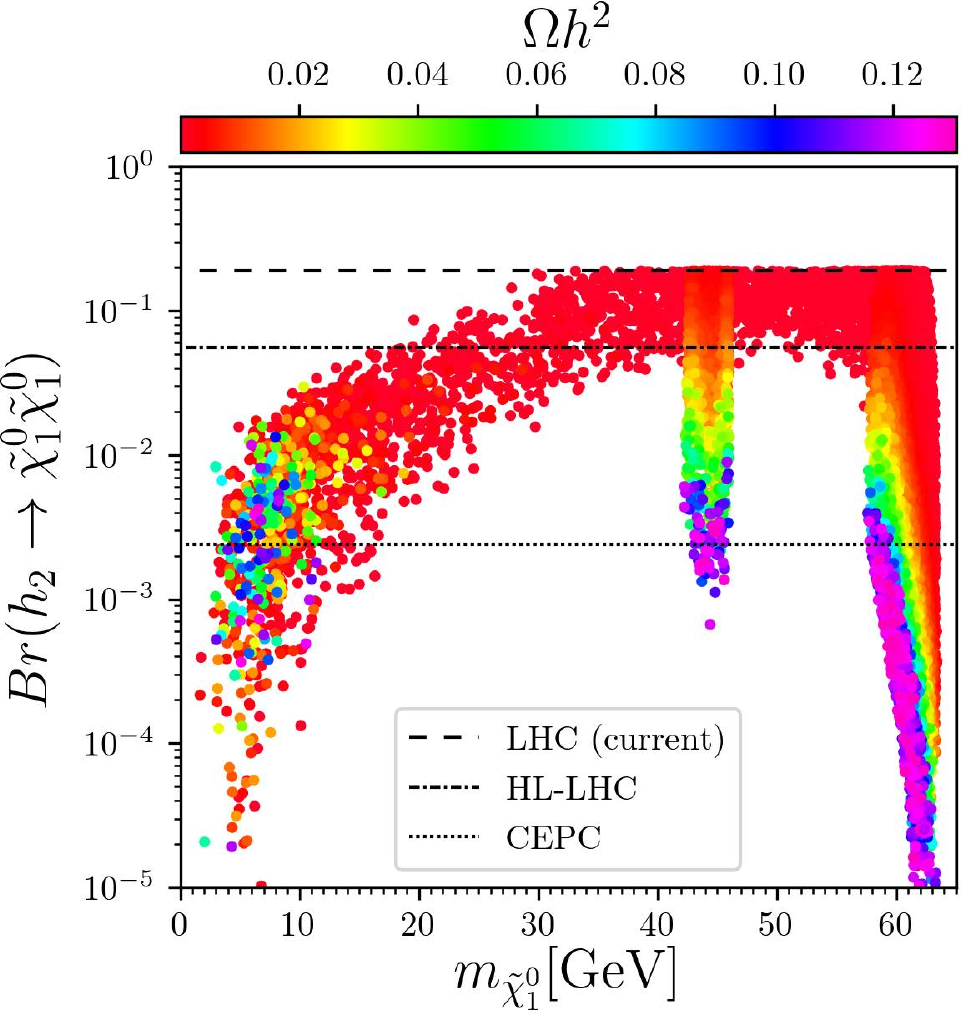}
\vspace{-0.6cm}
\caption{(color online). The surviving samples projected in the invisible branching ratio $Br(h_2\!\!\to\!\!\chi\chi)$ versus $m_{\chi}$ planes, with colors indicating the relic density $\Omega h^2$.
The dashed, dot-dashed and doted line show the current upper limit $19\%$ of the invisible Higgs decay \cite{Sirunyan:2018owy}, future detection accuracy of that at HL-LHC $5.6\%$ \cite{Liu:2016zki} and CEPC $0.24\%$ \cite{Tan:2020fxk} respectively.
The $\chi$ denotes the lightest neutralino $\tilde{\chi}^0_1$ in this paper.
}
\label{fig3}
\end{figure}
In Fig.\ref{fig3} we project the surviving samples in the Higgs invisible branching ratio $Br(h_2\to\chi\chi)$ versus $m_{\chi}$ planes, with colors indicate the relic density $\Omega h^2$.
From this figure, we can see that for $h_2$-funnel and $Z$-funnel annihilation of $\chi$ with right relic density, the invisible Higgs decay can be about $0.4\%$ and $1\%$ respectively at most.
While for $h_1$-funnel and $a_1$-funnel annihilation with right relic density, the invisible Higgs decay can be $2\%$ at most, and the corresponding $\chi$ mass is about $12\GeV$ at most.
All the samples with right relic density can not be detected at the HL-LHC, but may be covered at the CEPC.

\begin{table}[!htb]
\centering
\caption{\label{tab:benchmark}
One benchmark point for each of the four annihilation mechanisms: P1 for $h_2$-funnel, P2 for $Z$-funnel, P3 for $a_1$-funnel and P4 for $h_1$-funnel.
The theoretical cross section of $pp\to \chi^{\pm}_1 \chi^0_{2,3}$ are calculated with prospino 2 at NLO \cite{Beenakker:1999xh}.
While the upper limit $\sigma^{\rm lim}_{95\%}$ are taken from Fig.7 and 8(a) in Ref.\cite{Sirunyan:2018ubx}.
}
\footnotesize
\begin{tabular*}{80mm}{|c@{\extracolsep{\fill}}|c|c|c|c|}
\toprule
\hline
   ~~Point~~  & P1    & P2    & P3    & P4
\\ \hline
 $\lambda$
& $0.100$
& $0.0693$
& $0.161$
& $0.148$
\\ \hline
 $\kappa\; [10^{-2}]$
& $-2.05$
& $1.18$
& $0.182$
& $0.297$
\\ \hline
 $\tan\beta$
& 27.0
& 23.9
& 21.1
& 23.0
\\ \hline
 $\mu_{\rm eff}\;[{\rm GeV}]$
& 150
& 128
& 171
& 166
\\ \hline
 $A_{\lambda}\;[{\rm TeV}]$
& $3.75$
& $2.23$
& $2.98$
& $3.29$
\\ \hline
 $A_{\kappa}\;[{\rm GeV}]$
& 98.0
& 126
& 180
& 173
\\ \hline
 $M_0\;[{\rm GeV}]$
& 480
& 432
& 13.2
& 302
\\ \hline
 $M_{1\!/\!2}\;[{\rm TeV}]$
& 1.37
& 1.11
& 1.11
& 1.15
\\ \hline
 $A_0\;[{\rm TeV}]$
& -1.48
& -1.23
& -1.58
& -1.35
\\ \hline
 $m_{h_2}\;[{\rm GeV}]$
& 125.2
& 124.3
& 125.1
& 124.6
\\ \hline
 $m_{h_1}\;[{\rm GeV}]$
& 61.1
& 40.0
& 15.7
& 15.5
\\ \hline
 $m_{a_1}\;[{\rm GeV}]$
& 19.5
& 31.0
& 9.30
& 7.82
\\ \hline
 $m_\chi\;[{\rm GeV}]$
& 62.4
& 44.4
& 4.00
& 6.78
\\ \hline
 $m_{\chi^0_2}\;[{\rm GeV}]$
& 149
& 125
& 170
& 165
\\ \hline
 $m_{\chi^0_3}\;[{\rm GeV}]$
& 160
& 136
& 182
& 176
\\ \hline
 $m_{\chi^\pm_1}\;[{\rm GeV}]$
& 153
& 130
& 173
& 169
\\ \hline
 $m_{\tilde{\tau}_1}\;[{\rm GeV}]$
& 279
& 485
& 128
& 133
\\ \hline
 $m_{\tilde{\nu}_{\tau}}\;[{\rm GeV}]$
& 270
& 578
& 104
& 109
\\ \hline
 $\Omega h^2$ & 0.118 & 0.116 & 0.111 & 0.117 \\ \hline
 $~Br(h_2 \!\!\to\!\! \chi \chi)\;[10^{-3}]~$ & 0.03 & 3.94 & 2.19 & 3.52 \\ \hline
 $Br(h_1 \!\!\to\!\! \chi \chi)$&  0.000 & 0.000 & 0.244 & 0.117 \\ \hline
 $Br(a_1 \!\!\to\!\! \chi \chi)$&  0.000 & 0.000 & 0.194 & 0.000 \\ \hline
 $~Br(h_2 \!\!\to\!\! a_1 a_1)\;[10^{-3}]~$& 2.50 & 0.67 & 0.84 & 0.26 \\ \hline
 $~Br(h_2 \!\!\to\!\! h_1 h_1)\;[10^{-3}]~$& 0.13 & 0.52 & 2.08 & 0.45 \\ \hline
 $~Br(\chi^{\pm}_1 \!\!\to\!\! W^{\pm} \chi)~$ & 1.000 & 1.000 & 0.477 & 0.445 \\ \hline
 $~Br(\chi^{\pm}_1 \!\!\to\!\! \tilde{\nu}_{\tau} \tau^{\pm})~$ & 0.000 & 0.000 & 0.518 & 0.552 \\ \hline
 $~Br(\chi^{\pm}_1 \!\!\to\!\! \tilde{\tau}^{\pm}_1 \nu_{\tau})\;[10^{-3}]~$ & 0.00 & 0.00 & 4.89 & 2.81 \\ \hline
 $Br(\chi^0_2 \!\!\to\!\! Z \chi)$ & 0.000 & 0.000 & 0.322 & 0.333 \\ \hline
 $~Br(\chi^0_2 \!\!\to\!\! h_1 \chi)~$ & 0.028 & 0.710 & 0.021 & 0.019 \\ \hline
 $~Br(\chi^0_2 \!\!\to\!\! a_1 \chi)~$ & 0.612 & 0.085 & 0.012 & 0.011 \\ \hline
 $~Br(\chi^0_2 \!\!\to\!\! h_2 \chi)~$ & 0.000 & 0.000 & 0.141 & 0.144 \\ \hline
 $~Br(\chi^0_2 \!\!\to\!\! \tilde{\tau}_1^{\pm} \tau_1^{\mp})~$ & 0.000 & 0.000 & 0.388 & 0.387 \\ \hline
 $~Br(\chi^0_2 \!\!\to\!\! \tilde{\nu}_\tau \bar{\nu}_\tau)~$ & 0.000 & 0.000 & 0.117 & 0.106 \\ \hline
 $~Br(\chi^0_2 \!\!\to\!\! \chi f\bar{f})~$ & 0.360 & 0.205 & 0.000 & 0.000 \\ \hline
 $~Br(\chi^0_3 \!\!\to\!\! Z \chi)~$   & 0.957 & 0.306 & 0.422 & 0.419 \\ \hline
 $~Br(\chi^0_3 \!\!\to\!\! h_1 \chi)~$ & 0.030 & 0.142 & 0.013 & 0.010 \\ \hline
 $~Br(\chi^0_3 \!\!\to\!\! a_1 \chi)~$ & 0.014 & 0.552 & 0.012 & 0.011 \\ \hline
 $~Br(\chi^0_3 \!\!\to\!\! h_2 \chi)\;[10^{-3}]~$ & 0.000 & 0.000 & 0.096 & 0.080 \\ \hline
 $~Br(\chi^0_3 \!\!\to\!\! \tilde{\tau}_1^{\pm} \tau_1^{\mp})~$ & 0.000 & 0.000 & 0.420 & 0.443 \\ \hline
 $~Br(\chi^0_3 \!\!\to\!\! \tilde{\nu}_\tau \bar{\nu}_\tau)~$ & 0.000 & 0.000 & 0.037 & 0.036 \\ \hline
 $\sigma(pp\!\to\! \chi^{+}_1 \chi^0_2)\; [{\rm fb}]$ & 780 & 1410 & 496 & 548 \\ \hline
 $\sigma(pp\!\to\! \chi^{-}_1 \chi^0_2)\; [{\rm fb}]$ & 445 & 833 & 275 & 306 \\ \hline
 $\sigma(pp\!\to\! \chi^{+}_1 \chi^0_3)\; [{\rm fb}]$ & 683 & 1220 & 438 & 485 \\ \hline
 $\sigma(pp\!\to\! \chi^{-}_1 \chi^0_3)\; [{\rm fb}]$ & 387 & 714 & 241 & 269 \\ \hline
 $\sigma(pp\!\to\!W^{\pm}Z\chi\chi)\; [{\rm fb}]$ & 1481 & 1067 & 255 & 267 \\ \hline
 $\sigma^{\rm lim}_{95\%}\; [{\rm fb}]$ & 2957 & 1908 & 645 & 625 \\ \hline
\bottomrule
\end{tabular*}
\end{table}

In our calculating above, we employed \textsf{SModelS-v1.2.2} to impose the sparticle constraints, where the results of electroweakino searches especially in multi-lepton channels \cite{Sirunyan:2018ubx} constrained our scenario most since our higgsinos are light, but there are still lots of samples surviving at last with higgsinos at $100\!\!\sim\!\! 200\GeV$.
In Tab.\ref{tab:benchmark} we list one benchmark point for each of the four funnel channels, which can be used for further checking with update search results in the future.
As can be seen from the table, all the benchmark points predict correct relic density but small ratio of the invisible decay.
With the same method in Ref.\cite{Ellwanger:2018zxt}, we also calculate the rates of $pp\!\!\to\!\! \chi^{\pm}_1 \chi^0_{2,3} \!\!\to\!\! W^{\pm} Z\chi\chi$, verifying them under the $95\%$-CL upper limits in Fig.7 and 8(a) in Ref.\cite{Sirunyan:2018ubx}.
Moreover, one can see that for the $h_2/Z$-funnel points, $\chi^0_2$ cannot decay to on-shell $Z$ boson, but has a sizable branching ratio of three-body decay; while for the $h_1/a_1$-funnel points, the sleptons $\tilde{\tau}_1$ and $\tilde{\nu}_{\tau}$ can be lighter than higgsinos, thus higgsinos can have sizable branching ratios to a lighter slepton.
With the update \textsf{SModelS} (\textsf{-v1.2.3}), we also checked that these points meanwhile satisfy the constraints of stau search in hadronic-$\tau$ channels \cite{Aad:2019byo}, electroweakino searches in final states with tau leptons \cite{Sirunyan:2018vig}, in multi-lepton channels by Atlas with $139\fbm$ data \cite{Aad:2019vvi}, etc \cite{Sirunyan:2017zss, Aaboud:2018ngk, Aaboud:2018sua, Aaboud:2018jiw, Sirunyan:2018nwe, Aaboud:2018kya, Sirunyan:2018lul}.

\section{Conclusions}
\label{sec:conc}

In this work, we have discussed the funnel-annihilation mechanisms of light dark matter and the invisible Higgs decay in the scNMSSM, which is also called the NMSSM with non-universal Higgs masses (NUHM).
We use the scenario we got in our former work on scNMSSM, where we considered the theoretical constraints of vacuum stability and Landau pole, and experimental constraints of Higgs data, muon g-2, B physics, sparticle searches, dark matter relic density and direct searches, etc.
In our scenario, the bino and wino are heavy because of the high mass bound of gluino and the unification of gaugino masses at the GUT scale.
Thus the lightest neutralino $\chi$, or LSP, can only be singlino-dominated or higgsino-dominated.

Finally, we come to the following conclusions regarding the funnel-annihilation mechanisms of light dark matter and the invisible Higgs decay in scNMSSM:
\begin{itemize}
    \item There can be four funnel-annihilation mechanisms for the LSP $\chi$, which are the $h_2$, $Z$, $h_1$ and $a_1$ funnel.
    \item For the $h_1$ and $a_1$ funnel with right relic density, the $\chi$ mass is lighter than $12\GeV$, and the invisible Higgs decay can be $2\%$ at most.
    \item For the $h_2$ and $Z$ funnel with right relic density, the invisible Higgs decay can be about $0.4\%$ and $1\%$ respectively at most.
    \item If the invisible Higgs decay was discovered at the HL-LHC, the four funnel-annihilation mechanisms of light dark matter can be excluded with $\chi$ as the only dark matter source.
\end{itemize}

Moreover, it needs to emphasize that, the above conclusions are got in the semi-constrained NMSSM, where the parameters in the Higgs sector are all free, while these in the sfermion and gaugino sectors are constrained a lot.
Hence corresponding conclusions in the `unconstrained' NMSSM may be different, because when parameters in the sfermion and gaugino sectors are released, the surviving regions of parameters $\lambda$ and $\kappa$ can be enlarged, the singlet component in the Higgses can be enhanced, and the lightest neutralino $\chi$ can also be bino- or wino-dominated \cite{Cao:2012fz, Ellwanger:2011aa, King:2012is, Cao:2012yn, Cao:2013gba, Cao:2013mqa, Wang:2018vrr}.

\begin{acknowledgments}
\paragraph*{Acknowledgements.}
This work was supported by the National Natural Science Foundation of China (NNSFC)
under grant Nos. 11605123.
\end{acknowledgments}

\bibliography{ref}

\end{document}